# Comparing major declaration, attrition, migration, and completion in physics with other STEM disciplines: A sign of inequitable physics culture?

Kyle M. Whitcomb, Danny Doucette, Chandralekha Singh
University of Pittsburgh

*This research uses 10 years of institutional data at a large public university in the USA to investigate trends in the undergraduate majors students declare, drop, and earn degrees, especially comparing physics to other disciplines. We find that physics has the lowest number of students of all science, technology, engineering, and math (STEM) disciplines and it also has the highest rates of attrition of students who declare a major. While many STEM disciplines have students migrating both in and out of those majors, physics primarily has a uni-directional migration of students out of the major. Furthermore, physics has the lowest percentage of women undergraduate majors. Using an equity framework, we view these findings as signatures of inequitable and non-inclusive culture. We suggest that important roles may be played by stereotypes such as the incorrect belief that physics is accessible only to brilliant men, the issue of first-year college physics courses failing to energize students, and apathy in large physics departments toward improving intentional recruitment and retention of physics majors.*

*Keywords: attrition, gender, inequities, majors, retention*

**INTRODUCTION AND THEORETICAL FRAMEWORK**

Increasingly, Science, Technology, Engineering, and Mathematics (STEM) departments across the US are focusing on using evidence to improve the learning of all students and making learning environments equitable and inclusive (Whitcomb et al., 2020a, 2020b; Blue et al., 2018; Brewe et al., 2010; Hyater-Adams et al., 2018; Johnson, 2012; Johnson et al., 2017; Karim et al., 2018; King, 2016; Li and Singh, 2021; Little et al., 2019; Lorenzo et al., 2006; Maltese and Cooper, 2017; Maltese and Tai, 2011; McCavit and Zellner, 2016; Means et al., 2018; Metcalf et al., 2018; A. Traxler and Brewe, 2015; A. L. Traxler et al., 2016). However, women and racial and ethnic minority students continue to be severely underrepresented in many STEM disciplines such as physics (National Science Board, 2018; National Student Clearinghouse Research Center, 2015). Many prior studies have focused on identifying historical sources of inequities within society, i.e., societal norms that perpetuate obstacles to the participation and success of certain groups of disadvantaged people in education and beyond (Kalender et al., 2019; Crenshaw et al., 1995; Eddy and Brownell, 2016; Gonsalves et al., 2016; Gutiérrez, 2009; Henderson et al., 2017; Kalender et al., 2020; Kellner, 2003; Ladson-Billings and Tate, 1995; Metcalf et al., 2018; Rosa and Mensah, 2016; Schenkel and Calabrese Barton, 2020; Taylor et al., 2009; Tolbert et al., 2018; Yosso, 2005). These types of studies are very valuable because they provide a historical perspective on the inequities based upon gender, race, ethnicity etc. in STEM fields (Bang and Medin, 2010; Estrada et al., 2018; Ganley et al., 2018; Johnson, 2012; Johnson et al., 2017; Madsen et al., 2013; Metcalf et al., 2018; Ong et al., 2018; Schenkel and Calabrese Barton, 2020; Seron et al., 2016; Tolbert et al., 2018).

While prior studies have used both qualitative and quantitative approaches to investigate equity and inclusion in STEM, the use of institutional data to investigate past and current trends in enrollment in different majors, attrition from those majors, migration from one major to another and degree achievement can also shed light on equity and inclusion pertaining to different majors. In particular, in the past few decades, institutions have been keeping increasingly large digital databases of student records. Studies utilizing many years of institutional data can lead to analyses that were previously limited by statistical power. This is particularly true for studies of recruitment and retention in STEM majors such as physics that rely on large sample sizes (King, 2016; Maltese and Cooper, 2017; Maltese and Tai, 2011; Matz et al., 2017; Means et al., 2018; Safavian, 2019; Salehi et al., 2019; Shafer et al., 2021; Witherspoon and Schunn, 2019). We have now reached the point where there are sufficient data available at many institutions for analyses that can provide valuable information pertaining to certain aspects of equity and inclusion, e.g., recruitment and retention in different majors (Baker and Inventado, 2014; Papamitsiou and Economides, 2014).

In this study, we use 10 years of institutional data from a large state-related research university in the USA to investigate how patterns of female and male student major-declaration, dropping of the initial major and subsequent degree-earning in a major differ for those who initially declare physics major compared to other majors. Our investigation is motivated by the framework that physics continues to be one of the least diverse and the physics culture particularly at many large research universities with graduate programs is often apathetic to intentional recruitment of majors and making physics learning environment equitable and inclusive so that diverse groups of students can thrive as majors (Gonsalves et al., 2016; Henderson et al., 2017; Rosa and Mensah, 2016; Salehi et al., 2019). In particular, a lack of focus on equity and inclusion and supporting students with diverse backgrounds can lead to very few students declaring physics as their major. Those who declare it may drop out instead of earning a physics degree despite the fact that physics can be an intellectually stimulating and enjoyable subject that can help students become good problem solvers and critical thinkers, skills that would serve them well regardless of the type of career they eventually pursue.

Previously we investigated inequities in learning for ethnic and racial minority (ERM) students in STEM disciplines using 10 years' institutional data (Whitcomb et al., 2021). We found that ERM students drop most STEM majors at higher rates compared to other students. This is particularly true in physics in which the percentage of ERM students who dropped physics major is twice the percentage of White students who drop it (Whitcomb et al., 2021). These troubling trends signify systemic inequities, lack of student-centered pedagogy and sufficient support and mentoring for the ERM students, who are often already severely disadvantaged particularly at a predominantly White institution like ours.

Here we discuss an investigation at the same institution from the lens of lack of equity and inclusion based upon female and male students' recruitment, retention and graduation patterns in various majors with a focus on physics, i.e., our focus is particularly on how physics compares with other disciplines (Good et al., 2019). Within our framework, one mechanism by which societal stereotypes and biases about gender can influence student choice of major is proposed by Leslie et al., who showed that disciplines with a higher attribution of "brilliance" also have a lower representation of women (Leslie et al., 2015) due to pervasive stereotypes about men being "brilliant" in those disciplines. These brilliance-attributions affect all levels of STEM education, starting with early childhood when girls have already acquired these notions that girls are not as brilliant as boys (Bian, 2017; Bian et al., 2017), which can influence their interest in pursuing certain STEM disciplines (Bian, Leslie, Murphy, et al., 2018; Hazari et al., 2013; Ivie et al., 2016; Sax et al., 2016), and affect how likely they are to be referred for related opportunities (Bian, Leslie, and Cimpian, 2018).

Using the lens of equity, the goal of this study is to identify ways in which cultural factors associated with undergraduate studies in physics, impact the enrollment decisions made by female and male students. We interpret these results within an equity framework, which consists of three pillars: equitable opportunity and encouragement to learn within a major, equitable outcomes, and equitable and inclusive learning environments. By equity in learning, we mean that not only should all students have equitable opportunities and encouragement to participate in a major such as physics via intentional recruitment, they should also have an equitable and inclusive learning environment with appropriate support and mentoring so that they can engage in learning in a meaningful and enjoyable manner and the learning outcomes should be equitable. By equitable learning outcomes, we mean that students from all demographic groups (e.g., regardless of their gender identity or race/ethnicity) who have the pre-requisites to enroll in courses have comparable learning outcomes. This conceptualization of equitable outcome is consistent with the equity of parity model from Rodriguez et al. (2012). The learning outcomes include student retention in courses and major as a whole as well as positive evolution in their discipline-related motivational beliefs such as sense-of-belonging and self-efficacy because regardless of performance, students' motivational beliefs can influence their retention in their major. An equitable and inclusive learning environment should provide guidance, support and mentoring to all students as appropriate and ensure that students have high sense of belonging regardless of their prior preparation so long as they have the prerequisite basic knowledge and skills (Stewart et al., 2021). An equitable and inclusive learning environment would also ensure that all students embrace challenges as learning opportunity instead of being threatened by them and enjoy learning. We note that equitable access, encouragement and support to major in a discipline, equitable and inclusive learning environment and equitable outcomes are strongly entangled with each other. For example, if the learning environment is not equitable and inclusive in introductory to advanced physics courses, the learning outcomes are unlikely to be equitable and vulnerable students are unlikely to be retained.

Societal stereotypes and biases may affect all levels of STEM education, starting from early childhood (Bian et al., 2017). Students' perceptions of the field of physics and other STEM disciplines can influence their interest in pursuing studies in these STEM disciplines (DeWitt et al., 2019; Hazari et al., 2013) and can even affect how likely they are to be referred for related opportunities (Bian, Leslie, & Cimpian, 2018). The result is that women, as well as men from ERM groups, are less likely to complete undergraduate studies in many STEM fields (National Science Board, 2018). However, it is less clear what impact stereotypes and biases have on the rates at which students begin, drop, and transfer in and out of university STEM studies.

**RESEARCH QUESTIONS**

In this study, we use 10 years of institutional data from a large state-related research university in the USA to investigate how patterns of female and male student major-declaration, dropping of the initial major, and subsequent degree-earning in a major differ for those who initially declare physics major compared to other majors. A lack of focus on equity and inclusion and supporting students with diverse backgrounds can lead to very few students declaring physics as their major. Those who declare a major in physics may also drop out instead of earning a physics degree. Students may also change their major: transferring into a physics major, or transferring from physics into a different major. Therefore, this study focuses on trends in male and female students' declaration of a major, their retention within the physics major, and migration of students from one major to another. A major goal of this study is to compare these trends in physics with those in other disciplines. Specifically, in this paper, we seek to address the following research questions:

1. How many men and women major in each discipline?
2. How do the rates of attrition from the various majors differ? How do the rates of attrition from the various majors differ for men and women?
3. Among those students who drop a given major, what degree, if any, do those students earn? How do these trends differ for men and women?
4. What fraction of declared majors ultimately earn a degree in that major in each STEM subject area? How do these trends differ for men and women?

# METHODOLOGY

**Sample**

Using the Carnegie classification system (Indiana University Center for Postsecondary Research, 2018), the university at which this study was conducted is a public, high-research doctoral university, with balanced arts and sciences and professional schools, and a large, primarily residential undergraduate population that is full-time and reasonably selective with low transfer-in from other institutions.

The university provided for analysis the de-identified institutional data records of students with Institutional Review Board approval. In this study, we examined these records for $N = 18,319$ undergraduate students enrolled in two schools within the university: the School of Engineering and the School of Arts and Sciences. This sample of students includes all of those from ten cohorts who met several selection criteria, namely that the students had first enrolled at the university in a Fall semester, had provided the university with a self-reported gender, and the students had either graduated and earned a degree, or had not attended the university for at least a year as of Spring 2019. This sample of students is 49.9% female and had the following races/ethnicities: 77.7% White, 11.1% Asian, 6.8% Black, 2.5% Hispanic, and 2.0% other or multiracial.

**Measures**

*Gender*

In this study, we focus on female and male student trajectories as they progress towards their undergraduate degrees. We acknowledge that gender is not a binary construct; however, in self-reporting their gender to the university students were given the options of "male" or "female" and so those are the two self-reported gender categories that we are able to analyze. The student responses to this question were included in the institutional data provided by the university. Very few students opted not to provide information about gender, and so were not considered in this study. We used the answers of those students who chose either "male" ("M") or "female" ("F") to group students in order to calculate summary statistics on the measures described in this section.

*Declared Major and Degree Earned*

For each student, the data include their declared major(s) in each semester as well as the major(s) in which they earned a degree, if any. The majors were categorized as either STEM or non-STEM, with STEM courses being those courses taken from any of the following departments: biological sciences (including neuroscience), chemistry, computer science, any engineering department, geology and environmental science, mathematics (including statistics), and physics and astronomy. We note that for the purposes of this paper, "STEM" does not include the social sciences.

The data were transformed into a set of binary flags for each semester, one flag for each possible STEM major as well as specific flags for the non-STEM majors psychology and economics and a general non-STEM category for all other non-STEM majors. A similar set of flags was created for the degrees earned by students. From these flags, we tabulated a number of major-specific measures in each semester, including the current number of declared majors, the number of newly declared majors from the previous semester, the number of dropped majors from the previous semester, and the number of retained majors from the previous semester.

The total number of unique students that ever declared or dropped a major were also computed. The subset of students that dropped each major were further investigated and the major in which they ultimately earned a degree, if any, was determined.

Throughout this paper, we group the STEM majors into three clusters: biological sciences (including neuroscience); computer science and engineering; and mathematics (including statistics), chemistry, physics and astronomy, and geology and environmental science (collectively, mathematics and physical science). We additionally consider two non-STEM majors, economics and psychology, separately from the rest of the non-STEM majors. Economics and psychology are useful comparisons because of the mathematical requirements of economics and the proximity of psychology to the biological and natural sciences. Although psychology is sometimes included in STEM as a science, here we keep it distinct. When ordering majors (i.e., in figures and tables), the majors will be presented in the order they are listed in the previous two sentences. Note that

"engineering" groups together all engineering majors for departments in the School of Engineering at the studied university. These majors include chemical, computer, civil, electrical, environmental, industrial, and mechanical engineering as well as bioengineering and materials science.

We will make use of shortened labels for the majors in figures and tables. These shortened labels are defined in Table 1.

**TABLE 1**
**A LIST OF THE MAJORS CONSIDERED IN THIS STUDY AND THE SHORTENED LABELS USED TO REFER TO THOSE MAJORS IN TABLES AND FIGURES. NOTE THAT "ENGINEERING" IS A COMBINATION OF MANY ENGINEERING MAJORS OFFERED BY THE SCHOOL OF ENGINEERING.**

| Major | Short Label |
|---|---|
| Biological Sciences (including Neuroscience) | Bio |
| Computer Science | CS |
| Engineering | Engr |
| Mathematics (including Statistics) | Math |
| Chemistry | Chem |
| Physics and Astronomy | Phys |
| Geology and Environmental Science | Geo |
| Economics | Econ |
| Psychology | Psych |
| Other Non-STEM | Non-STEM |

*Year of Study*

The year in which the students took each course was calculated from the students' starting term and the term in which the course was taken. Since the sample only includes students who started in fall semesters, each "year" contains courses taken in the fall and subsequent spring semesters, with courses taken over the summer omitted from this analysis. For example, if a student first enrolled in Fall 2012, then their "first year" occurred during Fall 2012 and Spring 2013, their "second year" during Fall 2013 and Spring 2014, and so on in that fashion. If a student is missing both a fall and spring semester during a given year but subsequently returns to the university, the numbering of those post-hiatus years is reduced accordingly. If instead a student is only missing one semester during a given year, no corrections are made to the year numbering.

**Analysis**

Proportions of students in various groups (i.e., grouped by major and/or gender) are calculated along with the standard error of a proportion (Freedman et al., 2007). In particular, the proportions we report are

- the proportion of students in each major that are men or women,
- the proportion of men and women, respectively, that declare each subject as a major,
- the proportion of declared majors that drop the major,
- the proportion of those who drop each major that earn a degree in another major, and
- the proportion of all declared majors that ultimately earn a degree in that major.

All analyses were conducted using R (R Core Team, 2019), making use of the package tidyverse (Wickham, 2017) for data manipulation and plotting. Error bars on all plots are one standard error.

# RESULTS

With the goal of comparing the trends in physics to those in other STEM disciplines in the discussion section, we describe our results here for all STEM disciplines and identify noteworthy results in each analysis. In particular, the results section here describes the trends in different disciplines and the discussion section will focus on how the trends in physics relate to those in other disciplines.

## Major Declaration Patterns

There are many angles with which we can approach the first research question and investigate patterns of student major declaration. First, Fig. 1 shows the number of female and male students that ever declared each major separately. These results provide an important context for the upcoming analyses that may be partially explained by the number of students in each major.

**FIGURE 1**
**FOR EACH MAJOR ON THE HORIZONTAL AXIS, THE NUMBER OF UNIQUE STUDENTS IN THE SAMPLE THAT EVER DECLARED THAT MAJOR IS PLOTTED. SINCE STUDENTS MAY CHANGE MAJORS OR DECLARE MULTIPLE MAJORS, SOME STUDENTS MAY CONTRIBUTE TO THE COUNTS OF MORE THAN ONE MAJOR. THESE COUNTS ARE CALCULATED SEPARATELY FOR MEN AND WOMEN.**

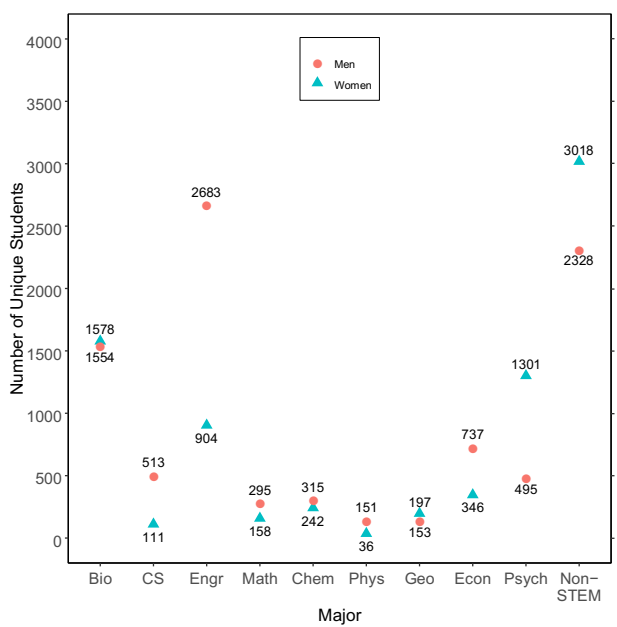

Figure 1 begins to hint at gender differences in enrollment patterns, such as a higher proportion of women majoring in non-STEM disciplines than men, or a higher proportion of men majoring in engineering than women. These gender patterns are explored further in Fig. 2 by standardizing the scales in two ways. In Fig. 2a, we consider the populations of each major separately and calculate the percentages of that population that are men or women. This provides insight into what these students might be encountering in the classes for their major. For instance, in the biological sciences there is a roughly even split, so students in biology classes for biological sciences majors might encounter a classroom that equally represents men and women. On the other end of the spectrum, around 80% of both computer science and physics and astronomy majors are men.

**FIGURE 2**
IN (A), THE PERCENTAGES OF STUDENTS IN EACH MAJOR THAT ARE MEN OR WOMEN ARE CALCULATED. A HORIZONTAL DASHED LINE OF SYMMETRY IS SHOWN AT 50%. IN (B), THE PERCENTAGES OF MEN AND WOMEN THAT MAJOR IN EACH SUBJECT ARE CALCULATED (I.E., THE PERCENTAGES FOR EACH GENDER GROUP WILL SUM TO ROUGHLY 100% IN THIS CASE). DISCREPANCIES IN THE SUM OF PERCENTAGES IN (B) MAY OCCUR DUE TO ROUNDING THE LISTED PERCENTAGES TO THE NEAREST INTEGER AS WELL AS STUDENTS DECLARING MULTIPLE MAJORS.

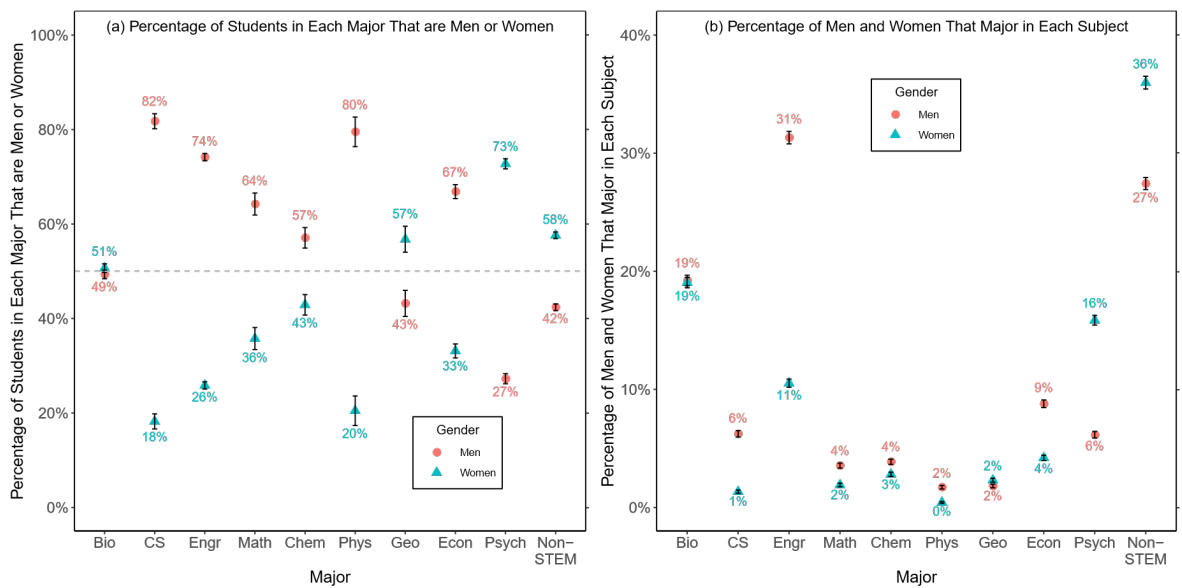

Another way to represent the population of these majors is to consider what percentage of all men or women choose each major, as seen in Fig. 2b. While this plot mimics that of Fig. 1, we can now read the differences noted earlier more clearly. In particular, the clearest differences in this view (Fig. 2b) are in engineering (31% of men and 11% of women declare an engineering major), non-STEM (27% of men and 36% of women declare a non-STEM major), and psychology (6% of men and 16% of women declare a psychology major).

Finally, we note that while we also considered the average term in which students declared each major, we did not find any notable gender trends in these results. We found that for almost all majors, the average term in which students declare the major is between the third and fourth terms (i.e., during the second year since students in the School of Arts and Sciences do not typically declare the major in their first year). There were two exceptions to this: engineering and computer science. The engineering departments are in the School of Engineering, separate from the School of Arts and Sciences at the studied institution, and thus students effectively declare an engineering major in their first term upon enrolling in the school. Meanwhile, the computer science department has very stringent requirements and does not allow students to declare the major until completion of five core courses within that major, which results in an average declaration of computer science major between the fifth and sixth terms.

These trends in engineering and computer science are important to keep in mind while considering the results presented later in this paper, since in computer science we are not able to capture attrition that occurs (of students intending to major) during the terms before a student officially declares a major. Conversely in engineering, we are able to capture almost all attrition in the first year due to the unique enrollment conditions of engineering students, which is not possible for majors within the School of Arts and Sciences (where students can declare their major at any time allowed by the department and it typically occurs in second year).

**Attrition Rates**

In order to answer the second research question, we further considered patterns of attrition rates by gender. In Fig. 3, we consider the drop rates of different subsets of students (all students, male students, and female students) in each major or group of majors. In Fig. 3a, we see that computer science, non-STEM, and psychology students are the least likely to drop their major, while physics, mathematics, and chemistry students are the most likely to drop. We note that the relatively low drop rate of computer science majors could be due to the late average declaration of the computer science major. That is, attrition from computer science prior to when students are allowed to declare the major is not accounted for in Fig. 3. The overall attrition rate in physics is the worst of all majors and shows that more than one third of the students drop the major after declaring it.

**FIGURE 3**
**FOR EACH MAJOR, THE PERCENTAGE OF STUDENTS WHO DECLARED THE MAJOR BUT SUBSEQUENTLY DROPPED THE MAJOR IS PLOTTED ALONG WITH ITS STANDARD ERROR. THIS IS DONE SEPARATELY FOR (A) ALL STUDENTS AND (B) MEN AND WOMEN SEPARATELY, ALONG WITH LINES CONNECTING DIFFERENT POINTS AS GUIDES TO THE EYE.**

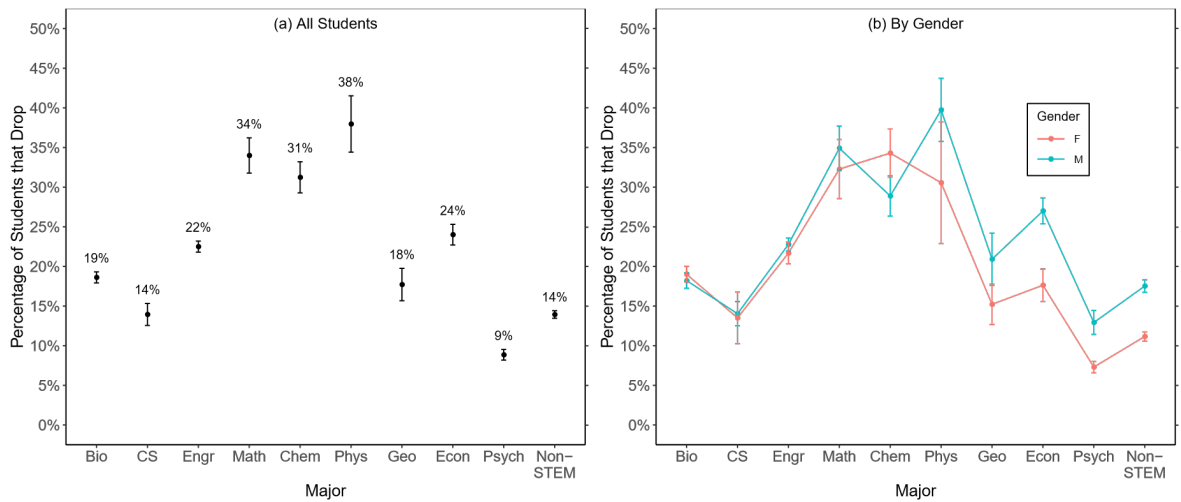

Fig. 3b shows the drop rates by gender, that is, separately for men and women. The drop patterns of men and women largely mimic the overall patterns in Fig. 3a, and while there are a few exceptions, the large error makes it difficult to draw any particular conclusions. Also, although we cannot claim any statistical significance, interviews with a few women (reported elsewhere (Doucette and Singh, 2020c)) in introductory physics courses hint at the possibility that they may be more likely to decide not to major in physics (even though they were planning to major in it when they arrived in their first year) after their experiences in their first-year introductory physics courses.

**Trajectories of Students After Dropping a Major**

After discussing how many students drop each major, we answer the third research question by plotting in Fig. 4 where those dropped majors ended up. In particular, the major indicated in the legends of Fig. 4a and 4b shows which major was dropped, while the plot shows the percentage of those who dropped that major and ultimately earned a degree in each of the majors on the horizontal axis, including the case when "no degree" was earned. For example, in Fig. 4a, we see that among the students that drop the physics major (indicated by the line color in the legend), roughly 15% of them end up earning a degree in mathematics (by looking at this line's value above "Math" on the horizontal axis). The figure also shows that the two most common destinations for those who drop any major are either no degree or a degree in non-STEM.

Apart from dropped STEM majors later earning degrees in non-STEM or leaving the university without a degree, we observe a few other interesting spikes. For instance, those who drop a physics major are likely to earn a degree in mathematics (Fig. 4a) and those who drop chemistry or physics (Fig. 4a) as well as biological science (Fig. 4b) are likely to earn engineering degrees. Further, those who drop from economics are likely to major in mathematics (Fig. 4b). While all students who drop any major are very likely to earn no degree, the percentage of dropped majors in this category exceeds 50% for computer science (Fig. 4a), non-STEM, and psychology.

**FIGURE 4**
**AMONG THE STUDENTS THAT DROP EACH STEM MAJOR AS WELL AS PSYCHOLOGY AND NON-STEM MAJORS, THE FRACTIONS OF STUDENTS THAT GO ON TO EARN A DEGREE IN OTHER MAJORS, OR WHO DO NOT EARN A DEGREE AT ALL, ARE PLOTTED ALONG WITH THEIR STANDARD ERROR. DROPPED MAJORS ARE GROUPED INTO (A) CHEMISTRY, COMPUTER SCIENCE, ENGINEERING, MATHEMATICS, AND PHYSICS AND ASTRONOMY MAJORS, AND (B) BIOLOGICAL SCIENCE, GEOLOGY, ECONOMICS, PSYCHOLOGY, AND NON-STEM MAJORS.**

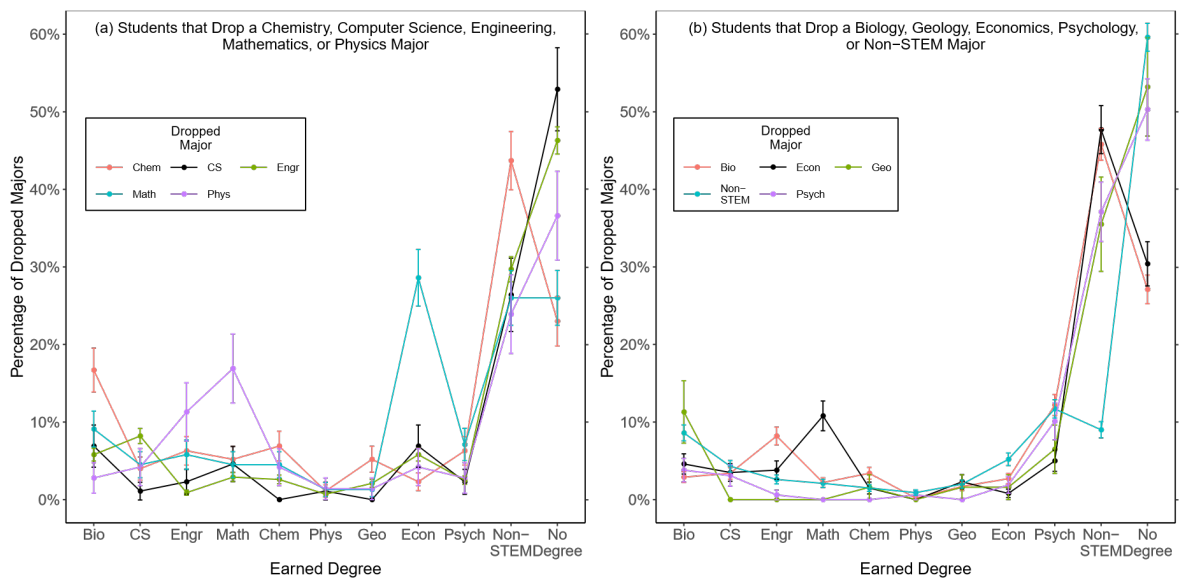

In order to further answer the third research question, Fig. 5 plots these same proportions of degrees earned by students who drop a major separately for men (Figs. 5a and 5b) and women (Figs. 5c and 5d). We observe for the most part very similar patterns between men and women, with a few notable differences. For example, among students who drop a chemistry degree, we observe that roughly 53% of the women eventually earn a degree in non-STEM (not including psychology or economics; Fig. 5c) compared with roughly 35% of the men (Fig. 5a). We observe a similar pattern with the roles reversed among those students who drop a biological sciences major, with roughly 15% of the men earning a degree in engineering (Fig. 5b) compared with less than 5% of the women (Fig. 5d). Another example is that men are more likely than women to earn computer science degrees after dropping a chemistry major (Figs. 5a and 5c). Finally, we note that across all of Fig. 5 in every major except psychology, the women who drop that major are either equally or more likely than the men to earn a degree in another major rather than leaving the university (that is, the women have a lower rate of earning "No Degree").

**FIGURE 5**
AMONG THE MEN AND WOMEN THAT DROP EACH STEM MAJOR AS WELL AS PSYCHOLOGY AND OTHER NON-SEM MAJORS, THE PERCENTAGES OF MEN AND WOMEN THAT GO ON TO EARN A DEGREE IN OTHER MAJORS, OR WHO DO NOT EARN A DEGREE AT ALL, ARE PLOTTED ALONG WITH THEIR STANDARD ERROR. DROPPED MAJORS ARE GROUPED INTO CHEMISTRY, COMPUTER SCIENCE, ENGINEERING, MATHEMATICS, AND PHYSICS AND ASTRONOMY MAJORS WHO ARE (A) MEN AND (C) WOMEN, AND BIOLOGICAL SCIENCE, GEOLOGY, ECONOMICS, AND PSYCHOLOGY MAJORS WHO ARE (B) MEN AND (D) WOMEN.

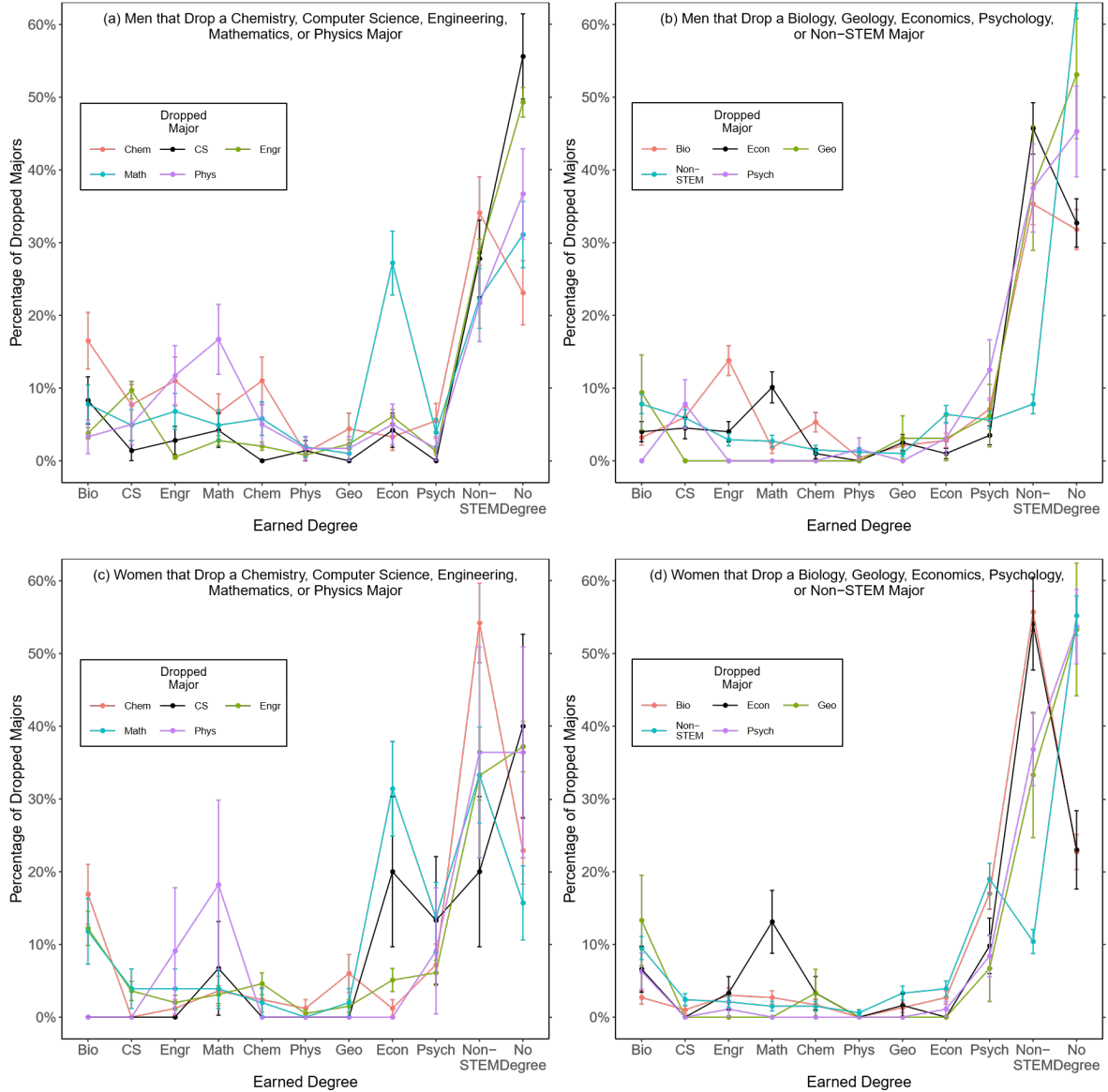

**Degree-Earning Rates**

In order to answer the fourth research question, we investigated how many students successfully earn a degree in each major. Figure 6a shows these degree-earning rates for all students in each major, while Fig. 6b shows these rates for female students and Fig. 6c for male students. While these are broadly similar to the reciprocal of the drop rates in Fig. 3, since some students drop a major and subsequently declare the same major again, these degree-earning rates are a more direct measurement of persistence in a major.

Looking first at the overall rates in Fig. 6a, there are fairly wide differences across majors, from the lowest rate in physics of about 65% to the highest in psychology and non-STEM, each at about 94%. The highest degree-earning rate in STEM occurs in computer science, with about 88% of declared computer science majors completing the degree requirements. As in Fig. 3, this can be at least partially explained by the requirements prior to declaring the major, which causes only students who have already progressed through a significant portion of the computer science curriculum to declare a computer science major.

Considering then the differences for women (Fig. 6b) and men (Fig. 6c), we see relatively few gender differences in these degree-earning rates. The slightly higher completion rate of women in non-STEM and psychology or men in chemistry, though statistically significant, are only differences of about 4-6%. As in Fig. 3, the largest difference between men and women seen here is in physics, with 75% of female physics majors earning a physics degree compared to 63% of male physics majors. However, the large error on these proportions, driven by the low sample size in physics shown in Fig. 1, makes it difficult to draw any conclusions from this gender difference in physics degree-earning rates. Similarly, women are more likely to complete a degree in economics, but again the size of the standard error prevents any conclusive statements about this difference.

Across all of Fig. 6, we note that since we have combined many majors for the "non-STEM" category, this is only a measure of the number of non-STEM majors who successfully earn a degree in any non-STEM major. That is, a student who drops one non-STEM major but earns a degree in a different non-STEM major will still be counted as having successfully earned a non-STEM degree. The same is true for the "engineering" category which also combine several majors. The high "success rates" of computer science and psychology may be due in part to the structure of their program encouraging students to declare somewhat later than other disciplines, and so this measure may not be capturing attrition that happens prior to an official declaration of major (e.g., a student intending to major in a discipline decides against it before ever declaring that major). On the other hand, since all students enrolled in the engineering school are considered to be majoring in "engineering" from the very beginning, the relatively low degree-earning rate of engineering reflects attrition even from the first to the second term, which is not captured for many other majors in which most students have not yet formally declared a major in their first term. Thus, each reported degree-earning rate here is a ceiling on the true rate that would include those students who intended to major but never declared.

**FIGURE 6**
**FOR EACH MAJOR LISTED ON THE HORIZONTAL AXIS, THE PERCENTAGES OF (A) ALL STUDENTS, (B) FEMALE STUDENTS, AND (C) MALE STUDENTS WHO DECLARE THAT MAJOR AND THEN EARN A DEGREE IN THAT MAJOR ARE PLOTTED ALONG WITH THE STANDARD ERROR.**

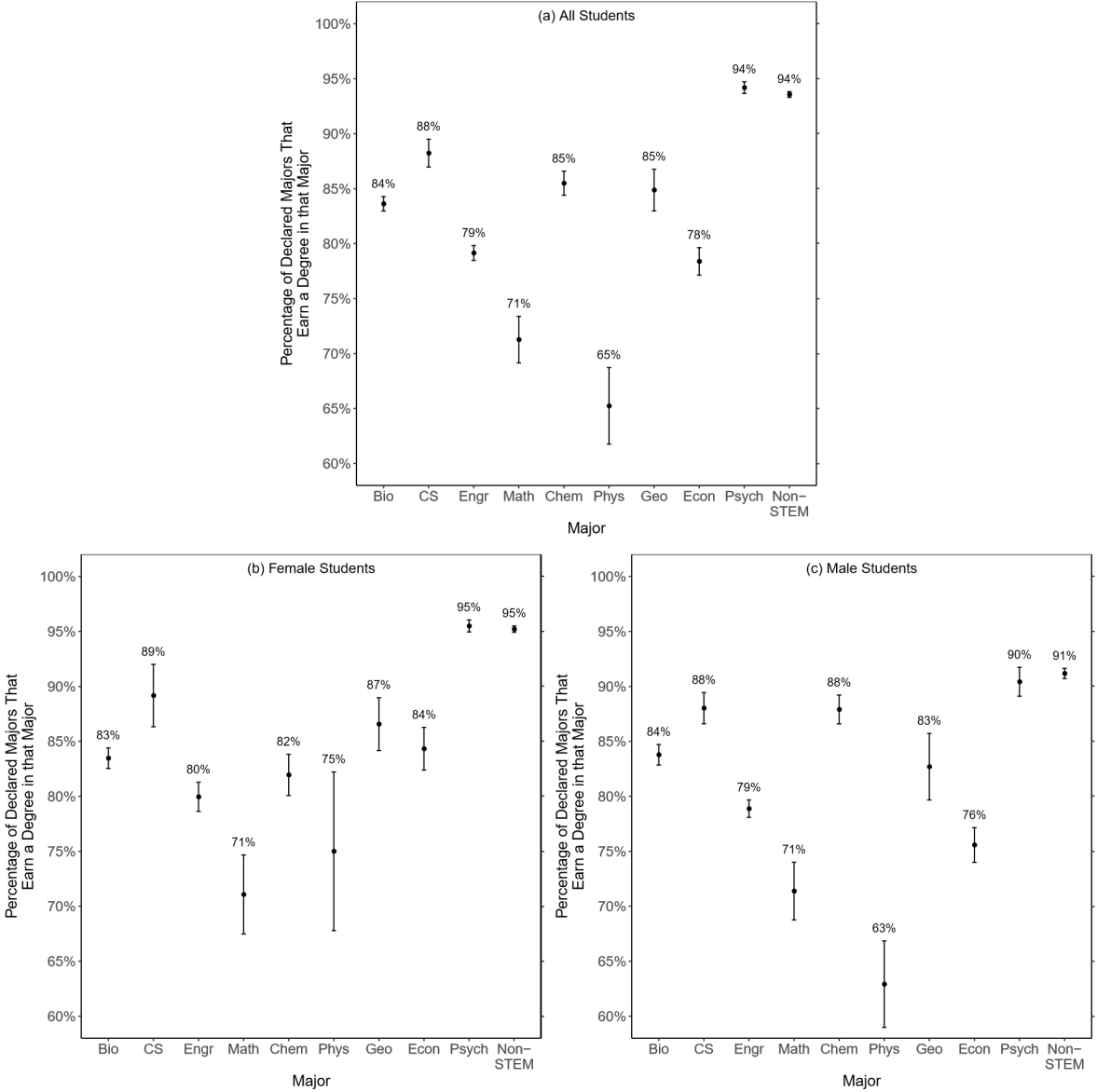

# DISCUSSION

In this section, we discuss the physics trends compared to other STEM disciplines starting with a discussion of the general trends (i.e., setting aside the gender differences), and then follow up with a discussion of the gender differences.

**General Enrollment Patterns**

We first note that physics has the lowest enrollment of all STEM disciplines (see Fig. 1). Moreover, despite large differences in the number of students enrolling in different STEM disciplines at the studied university (Fig. 1), there are broadly similar patterns of when those students declare the major (an average between the third and fourth term), with some exceptions (i.e., engineering and computer science due to the constraints on when a student can declare a major). The low enrollment in physics is consistent with the Leslie et al. (2015) study which identifies physics as a STEM discipline with the highest "ability belief" (i.e., emphasis on brilliance). Within our framework, when students encounter these types of societal stereotypes about brilliance, they are likely to feel that the physics major is not feasible for them even though it is an exciting field and an interesting way to develop their critical thinking and problem-solving skills.

Moreover, there are notable differences in the attrition of students, e.g., from a major such as physics and many other majors (Fig. 3), as well as the corresponding degree-earning rates (Fig. 6a). Notably, a few STEM disciplines, e.g., physics and mathematics, stand out as having particularly high rates of attrition (or low rates of degree completion) for students who declared those majors. This trend of high attrition rates is particularly problematic for physics (which has the highest attrition rate of all STEM disciplines), which recruits very few students in the first place (Fig. 1) with women comprising only 20% of those students (Fig. 2). Many students who declare a physics major drop out and move to other majors such as mathematics but students do not generally migrate to physics after declaring mathematics or other majors. Thus, physics is the worst major in terms of having only an outward flow of students from the discipline after they declare physics as their major.

Within our framework, these trends point to the absence of a culture that values students' needs and competencies. In particular, if physics learning environments are not student-centered, supportive, equitable and inclusive, students are likely to leave the discipline after declaring the major as observed in this investigation. The physics culture at a large research university can be particularly harmful in this regard. In physics, students who only have undergraduate degree in physics are not considered physicists unlike chemistry or engineering and many physics professors think of the task of supporting, mentoring, and educating students who are not planning to go to graduate school as an unproductive use of their time. This type of exclusive mentality can lead to students who are interested in majoring in physics but are not interested in graduate degree in physics being marginalized and wanting to drop the major after declaring it. The culture of physics particularly in large physics departments with graduate programs discussed here often entails lack of adequate support and mentoring for undergraduate majors based upon the assumption that if students were capable, they would figure out how to thrive on their own and if not, they are free to leave the major. This type of deficit thinking by physics faculty can drive students out of the major even if they wanted to major in physics and would have thrived with appropriate support and mentoring.

**Gendered Enrollment Patterns**

The most notable example of gender differences in enrollment patterns observed in our analysis is in Fig. 2. In biological science, chemistry and geological and environmental science, we see a more balanced representation of men and women. However, we see an underrepresentation of women in computer science, engineering, mathematics, physics, and economics, and a corresponding overrepresentation of women in non-STEM including psychology (Fig. 2a). Fig. 2 shows that physics and computer science have the lowest representation of women. Again, within our framework, these results are consistent with the fact that disciplines such as physics that have stereotypes about brilliance have low enrollment of women (Leslie et al., 2015). The gender imbalance in these STEM disciplines such as physics can itself play a pernicious role in recruitment and retention of women who do not have many role models and who are constantly forced to prove themselves and counter the societal stereotypes working against them. The issue can be especially salient for women considering physics as a major, since physics has one of the worst stereotypes about requiring a high degree of innate ability (Leslie et al., 2015).

We note that the attrition rates for women and men who declare a STEM major appear to be comparable in many STEM disciplines such as physics (see Fig. 3). However, individual interviews with a few female students in introductory physics courses in their first year suggest that they decided not to major in physics even though they were originally planning to major in it when they came to college (Doucette et al,, 2020c). Since students declare a physics major mainly in their second year, it is impossible to obtain a quantitative account of the possible differential attrition of female and male students intending to major in physics before they even officially declared their major using institutional data. However, it will be useful in the future to poll students at the beginning of their first year physics courses to investigate whether there is a differential attrition of students by gender who were intending to major in physics (even before declaring the major) due to the first year introductory physics courses more negatively impacting women than men (Marshman et al., 2018; Johnson et al., 2017; Matz et al., 2017).

**Trajectories of Students Who Drop a Major**

As with the attrition and degree-earning rates, we see broadly similar patterns between men and women who drop the various STEM majors (Fig. 5). As alluded to earlier, one noteworthy finding here is that the attrition of students in physics is primarily uni-directional; i.e., while we see a higher percentage of students leaving physics compared to other disciplines (Fig. 3), we also see very few students who dropped another major choosing to pursue physics (Figs. 4).

**CONCLUSION AND IMPLICATIONS**

Our findings using 10 years of institutional data at a large research university suggest that physics has the lowest number of majors and the lowest percentage of women (along with computer science) even though physics major can potentially be intellectually rewarding and help students develop critical thinking skills useful for a variety of careers. Within our framework, we posit that the stereotypes surrounding physics and brilliance as well as a physics culture at many large research universities that does not value supporting and mentoring undergraduates who may not be interested in graduate studies in physics may be partly responsible for these findings. Increasing the number of physics majors in general, and women majors in particular, would require intentional recruitment and retention efforts that center student assets and focus on addressing cultural factors such as biases and stereotypes.

Using our framework of equity, our findings for physics having the highest rates of attrition indicate that not only is there need to improve the support for the intended majors but also to make the physics learning environments equitable and inclusive. Focus on increasing equity and inclusion in learning is especially important in the early courses, since they are fraught with problematic gender differences and may partly be contributing to the underrepresentation of women in these majors in the first place. Our prior research suggests that after introductory physics, there was no gender difference in advanced physics course performance but there was in introductory courses but introductory physics course performance which had gender differences did not predict future course performance (Whitcomb et al., 2020a). Focusing on making introductory physics courses equitable and inclusive could also lead to an increase in the number of women wanting to major in physics because students declare a major in their second year. Moreover, improved performance as well as higher self-efficacy (Marshman et al., 2017, 2018) in student-centered introductory physics courses have the potential to motivate more women to choose a physics major. Similar steps should also be taken for the other STEM majors that have low representation of women, especially computer science and engineering.

All of these issues including improving the culture must be addressed urgently since they are critical for improving equity and inclusion in higher education learning environments in physics and related disciplines. Faculty members in physics and related disciplines should be provided incentives and support to accomplish these goals. They should be provided individual and collective opportunities with peers (e.g., at faculty retreats) to reflect upon the fact that intentional recruitment and retention of physics majors with varying interests with regard to career outcomes should be an important departmental goal even for a department with a large graduate program. Otherwise, the current status quo will persist and many students, who will excel with appropriate mentoring and support, will continue to be deprived of the opportunity and benefits of pursuing a major in physics and related disciplines.

**FUTURE DIRECTIONS**

This study was conducted at one large research university in the US and our data highlight important issues regarding equity in the declaration of physics major and subsequent attrition. This is true both in general, where we see physics having the highest rates of attrition among all STEM majors, and with regard to gender, where we see physics tied with computer science for the most unbalanced discipline by gender. Thus, an extension of this work would be for other institutions of different types and sizes in different countries to conduct similar studies in order to understand how widely applicable these findings are and continue to work towards the goal of equity in learning and creating equitable and inclusive learning environments in physics and other disciplines with inequitable trends. Other institutions noting similar highly problematic trends can help pinpoint common sources of inequities, while institutions that do not observe these trends may be able to identify how they have structured their student-centered programs to avoid these inequitable trends so that others can learn from them. We hypothesize that large research universities in the US similar to ours are likely to find similar trends if they do not have intentional focus on equity and inclusion and centering students in their instructional design. However, primarily undergraduate institutions of different types may have different trends with regard to female and male students' enrollment, attrition and degree attainment. These similarities and differences could provide useful insight into productive approaches to improve student recruitment and retention in physics and other disciplines with problematic trends.

Finally, this study focuses on female and male students' declaration of majors, attrition of students from those majors and degree attainment with a focus on physics. Our prior investigation focused on similar issues for ethnic and racial minority students at the same institution (Whitcomb et al., 2021). Future studies with similar types of institutional data sets could investigate intersectionality issues, e.g., inequities for ERM women, in order to investigate how under-served populations with multiple disadvantages and marginalized identities are being served. However, such study would require collection of data over a longer term, or pooling data from multiple institutions, in order to have statistical power.